\documentclass[11pt]{article}
\usepackage{titling}
\usepackage[OT1]{fontenc}
\usepackage{amsmath}
\usepackage{mathrsfs}
\usepackage{amsfonts}
\usepackage{amssymb}
\usepackage{diagbox}
\usepackage{footnote}
\usepackage{graphicx}
\usepackage{float}
\usepackage{epstopdf}
\usepackage[margin=1.25in]{geometry}
\usepackage{setspace}
\usepackage{tikz}
\usepackage{pgfplots}
\usepackage[round]{natbib}
\usepackage{multirow}
\usepackage[toc,page]{appendix}
\usepackage{hyperref}
\usepackage{caption}
\usepackage{subcaption}
\usepackage{enumitem}
\usepackage{booktabs}
\usepackage{chapterbib}
\usepackage{algorithm}
\usepackage{titletoc}
\providecommand{\U}[1]{\protect\rule{.1in}{.1in}}

\hypersetup{
	bookmarks=true,
	colorlinks=true,
	linkcolor=blue,
	citecolor=blue,
	filecolor=magenta,
	urlcolor=blue,
 breaklinks=true,
}

\newtheorem{assumption}{Assumption}[section]

\newtheorem{lemma}{Lemma}[section]

\newtheorem{proposition}{Proposition}[section]
\newtheorem{remark}{Remark}[section]

\begin{document}

    \title{Testing Partial Instrument Monotonicity}
	\author{ Hongyi Jiang \\China Center for Economic Research\\National School of Development\\ Peking University\\ hyjiang2017@nsd.pku.edu.cn\\
	\and Zhenting Sun\thanks{This work was supported by the National Natural Science Foundation of China [Grant Number 72103004].}\\China Center for Economic Research\\National School of Development\\ Peking University\\
	zhentingsun@nsd.pku.edu.cn}

	\maketitle

	\begin{abstract}
		When multi-dimensional instruments are used to identify and estimate causal effects, the monotonicity condition may not hold due to heterogeneity in the population. Under a partial monotonicity condition, which only requires the monotonicity to hold for each instrument separately holding all the other instruments fixed, the 2SLS estimand can still be a positively weighted average of LATEs. In this paper, we provide a simple nonparametric test for partial instrument monotonicity. We demonstrate the good finite sample properties of the test through Monte Carlo simulations. We then apply the test to monetary incentives and distance from results centers as instruments for the knowledge of HIV status.

	\end{abstract}
	
	\textbf{Keywords:} Partial monotonicity, instrument validity, nonparametric test

	\newpage

    \section{Introduction}
    
Exclusion, random assignment, and monotonicity are three fundamental conditions for instrument variables (IVs) to be valid in the identification and estimation of causal effects. When the instrument variable is multi-dimensional, \citet{mogstad2021causal} show that the monotonicity condition only holds if choice behavior is effectively homogeneous. However, such homogeneity may break down in many empirical applications. 
    \citet{mogstad2021causal} then consider a weaker version of monotonicity, partial monotonicity, which permits heterogeneous
choice behavior. Under partial monotonicity, multi-dimensional instruments (multiple instruments) can be used for causal effects estimation even if the full monotonicity condition fails.

    In the literature, the IV validity conditions can be tested by the methods of \citet{huber2015testing}, \citet{kitagawa2015test},  \citet{mourifie2016testing}, and \citet{sun2018ivvalidity}. In this paper, we extend the frameworks of \citet{kitagawa2015test} and \citet{sun2018ivvalidity} to the partial monotonicity of \citet{mogstad2021causal}. We show that the proposed test based on \citet{kitagawa2015test} and \citet{sun2018ivvalidity} performs well in practice. In the appendix, Monte Carlo studies demonstrate the finite sample properties of the test. We then apply the test to an empirical example from \citet{thornton2008demand} discussed in \citet{mogstad2021causal}.

 
	\section{Setup and Test Formulation}
	
	
	\citet{mogstad2021causal} introduce the concept of partial instrument monotonicity for multi-dimensional IVs. We follow \citet{mogstad2021causal} and mainly focus on the multivalued ordered treatment case.\footnote{It would be easy to extend the test to unordered treatments.} Let $( \Omega, \mathcal{A}, \mathbb{P} )$ be a probability space on which all the random elements are well defined. Suppose that the outcome variable $Y\in\mathbb{R}$, the treatment $D\in\mathcal{D}=\left\{
	d_{1},\ldots,d_J\right\}$ for some $J\ge 2$, and the instrument $\boldsymbol{Z}\in\mathcal{Z}=\{\boldsymbol{z}_1,\ldots,\boldsymbol{z}_K\}$ for some $K\ge 2$, where  every value $\boldsymbol{z}\in\mathcal{Z}$ is a vector $\boldsymbol{z}=(z_1,\ldots,z_L)$ for some $L\ge 2$. That is, $\boldsymbol{Z}$ is an $L$-dimensional instrument with $\boldsymbol{Z}=(Z_1,\ldots,Z_L)$, where $Z_l$ is a scalar variable for every $l$. Suppose the $l$th dimension of $\boldsymbol{Z}$ has $k_l$ possible values, i.e., $Z_l\in\{z_l^1,\ldots,z_l^{k_l}\}$. Following the rectangular support assumption of \citet{mogstad2021causal}, we suppose that $\mathcal{Z}=\mathrm{supp}(Z_1)\times\cdots\times\mathrm{supp}(Z_L)$, which implies $K=\prod_{l=1}^L k_l$. For every $\boldsymbol{z}=(z_1,\ldots,z_L)$, we define the vector $z_{-l}=(z_1,\ldots,z_{l-1},z_{l+1},\ldots,z_L)$, and $\boldsymbol{z}$ may be written as $\boldsymbol{z}=(z_{l},z_{-l})$. 
	Suppose that $Y_{d\boldsymbol{z}}$
	for $d\in\mathcal{D}$ and $\boldsymbol{z}\in\mathcal{Z}$, and  $D_{\boldsymbol{z}}$ for $\boldsymbol{z}\in
	\mathcal{Z}$ are the potential random variables. The following assumption formalizes the IV validity assumption for multi-dimensional
	instrument $\boldsymbol{Z}$ proposed by \citet{mogstad2021causal}.
	
	\begin{assumption}
		\label{ass.IV validity for multivalued Z}Partial IV validity for multi-dimensional $\boldsymbol{Z}$:
		\begin{enumerate}[label=(\roman*)]
			
			\item Instrument Exclusion: Almost surely, $Y_{d\boldsymbol{z}_{1}}=\cdots=Y_{d\boldsymbol{z}_{K}}$ for all $d\in\mathcal{D}$.
			
			\item Random Assignment: The variable $\boldsymbol{Z}$ is jointly independent of $(
			\tilde{Y},\tilde{D})  $, where
			\begin{align*}
				\tilde{Y}   =\left(  Y_{d_{1}\boldsymbol{z}_{1}},\ldots,Y_{d_{1}\boldsymbol{z}_{K}},\ldots, Y_{d_{J}\boldsymbol{z}_{1}
			},\ldots,Y_{d_{J}\boldsymbol{z}_{K}}\right) \text{ and }\tilde{D} =\left(  D_{\boldsymbol{z}_{1}},\ldots,D_{\boldsymbol{z}_{K}}\right)  .
			\end{align*}

			\item Partial Instrument Monotonicity: For every $l\in\{1,\ldots,L\}$ and every given $z_{-l}$, the potential treatment response variables satisfy $D_{(z_{l}^{k+1},z_{-l})}\geq D_{(z_{l}^{k},z_{-l})}$ almost surely for all
			$k\in \{1,\ldots, k_{l}-1\}$.
			
		\end{enumerate}
	\end{assumption}

\begin{remark}
Assumption \ref{ass.IV validity for multivalued Z}(iii) is a simplified version of Assumption PM (partial monotonicity) in \citet{mogstad2021causal}. \citet{mogstad2021causal} assume no direction in Assumption PM: For every $l\in\{1,\ldots,L\}$ and every given $z_{-l}$, the potential treatment response variables satisfy $D_{(z'_{l},z_{-l})}\geq D_{(z_{l},z_{-l})}$ almost surely or $D_{(z'_{l},z_{-l})}\leq D_{(z_{l},z_{-l})}$ almost surely for all
			$z_{l},z'_{l}\in \{z_l^1,\ldots,z_l^{k_l}\}$.
In practice, we may usually assume a direction in Assumption PM: For every $l$ and every $z_{-l}$, there is a sequence $\{z_{l}^{1}(z_{-l}),\ldots,z_l^{k_l}(z_{-l})\}$ such that 
\begin{align}\label{eq.generalized monotonicity}
D_{(z_{l}^{k+1}(z_{-l}),z_{-l})}\geq D_{(z_{l}^{k}(z_{-l}),z_{-l})}    
\end{align}
almost surely for all
			$k\in \{1,\ldots, k_{l}-1\}$.
For notational simplicity, we construct the test based on Assumption \ref{ass.IV validity for multivalued Z}(iii) which assumes that $z_l^k(z_{-l})=z_l^k$ for all $z_{-l}$. It would be straightforward to extend the test to \eqref{eq.generalized monotonicity}. We may also extend the test to Assumption PM in \citet{mogstad2021causal} following an idea similar to that in \citet[Section D]{sun2018ivvalidity} if there is no prior information about the direction in Assumption PM.
\end{remark} 

Suppose that $D$ has maximum value $d_{\max}$ and minimum value $d_{\min}$. We provide a testable implication for Assumption \ref{ass.IV validity for multivalued Z} in the following lemma.
	
	\begin{lemma}
		\label{lemma.testable implication}
  For all $1 \leq l \leq L$, all $1\le k \le k_{l}-1$, all possible values $z_{-l}$,  all Borel sets $B$, and all $C=(-\infty,c]$ with $c\in\mathbb{R}$, it follows that
		\begin{align}\label{eq.testable implication multivalue}
			&\mathbb{P}\left(  Y\in B,D=d_{\max}|\boldsymbol{Z}=(z_{l}^{k},z_{-l})\right)     \leq \mathbb{P}\left(  Y\in B,D=d_{\max}|\boldsymbol{Z}=(z_{l}^{k+1},z_{-l})\right)\nonumber\\
			&\text{and }\mathbb{P}\left(  Y\in B,D=d_{\min}|\boldsymbol{Z}=(z_{l}^{k},z_{-l})\right)     \geq \mathbb{P}\left(  Y\in B,D=d_{\min}|\boldsymbol{Z}=(z_{l}^{k+1},z_{-l})\right); \\
			&\mathbb{P}\left(  D\in C|\boldsymbol{Z}=(z_{l}^{k},z_{-l})\right)     \geq \mathbb{P}\left(  D\in C|\boldsymbol{Z}=(z_{l}^{k+1},z_{-l})\right). \label{eq.fosd multi}
		\end{align}
	\end{lemma}

Lemma \ref{lemma.testable implication} can be proved analogously to Lemma 2.1 of \citet{sun2018ivvalidity}. 	
In the following, we extend the tests of \citet{kitagawa2015test} and \citet{sun2018ivvalidity} to partial IV validity.
Without loss of generality, we assume that $d_{\min}=d_1\le\cdots\le d_J=d_{\max}$ with $d_{\min}=0$ and $d_{\max}=1$. Then the inequalities in \eqref{eq.testable implication multivalue} and \eqref{eq.fosd multi} are equivalent to 
	\begin{align}\label{eq.testable implication inequalities multi}
		&(-1)^{d}\cdot\left\{\mathbb{P}\left(  Y\in B,D=d|\boldsymbol{Z}=(z_{l}^{k+1},z_{-l})\right)-\mathbb{P}\left(  Y\in B,D=d|\boldsymbol{Z}=(z_{l}^{k},z_{-l})\right)\right\}  \leq 0  \notag\\
		&\text{ and } \mathbb{P}\left( D\in C|\boldsymbol{Z}=(z_{l}^{k+1},z_{-l})\right)-\mathbb{P}\left( D\in C|\boldsymbol{Z}=(z_{l}^{k},z_{-l})\right)\le 0
	\end{align}
	for all $1 \leq l \leq L$, all $1\le k\le k_{l}-1$, all possible values $z_{-l}$, all closed intervals $B$ in $\mathbb{R}$, each $d\in\{0,1\}$, and all $C=(-\infty,c]$ with $c\in\mathbb{R}$. Following {Lemma B.7} of \citet{kitagawa2015test}, here we only consider all closed intervals $B$ instead of all Borel sets $B$ when constructing the test. 
	By definition, for all $B,C\in\mathcal{B}_{\mathbb{R}}$ and all possible values $\boldsymbol{z}$,
	$
	\mathbb{P}\left(  Y\in B,D\in C|\boldsymbol{Z}=\boldsymbol{z}\right)={\mathbb{P}\left(  Y\in B,D\in C,\boldsymbol{Z}=\boldsymbol{z}\right)
	}/{\mathbb{P}\left(  \boldsymbol{Z}=\boldsymbol{z}\right)  }.
	$
	We then define function spaces
	\begin{align}{\label{def.function spaces}}
		&\mathcal{G}^l_{z_{-l}}=\left\{  \left(  1_{\mathbb{R}\times\mathbb{R}  \times\{  (z_{l}^{k},z_{-l})\}  },1_{\mathbb{R}\times\mathbb{R}  \times\{  (z_{l}^{k+1},z_{-l})\}  }\right)
		:k=1,\ldots,k_l-1\right\}\text{ for every }l \text{ and every }z_{-l}, \notag \\
        &{\mathcal{G}}=\cup_{l}\cup_{z_{-l}}\mathcal{G}^l_{z_{-l}},\notag\\
		&\mathcal{H}_{1}=\left\{  \left(  -1\right)  ^{d}\cdot1_{B\times\left\{
			d\right\}  \times\mathbb{R}^L}:B\text{ is a closed interval in }\mathbb{R},
		d\in\{0,1\}\right\},\notag\\
		&\bar{\mathcal{H}}_1=\left\{  \left(  -1\right)  ^{d}\cdot1_{B\times\left\{
			d\right\}\times\mathbb{R}^L  }:B\text{ is a closed, open, or half-closed interval in }
		\mathbb{R}
		,d\in\left\{  0,1\right\}  \right\},\notag\\
		&\mathcal{H}_{2}=\left\{  1_{\mathbb{R}\times C \times\mathbb{R}^L}:C=(-\infty,c],c\in\mathbb{R}\right\},\notag\\
		& \bar{\mathcal{H}}_2=\left\{  1_{\mathbb{R}\times C \times\mathbb{R}^L }: C=(-\infty,c] \text{ or } C=(-\infty,c),c\in\mathbb{R}  \right\},\notag\\
		&\mathcal{H}=\mathcal{H}_{1}\cup\mathcal{H}_{2}, \text{ and } \bar{\mathcal{H}}=\bar{\mathcal{H}}_1\cup\bar{\mathcal{H}}_2.
	\end{align}

    Let $\{\left(  Y_i,D_i,\boldsymbol{Z}_i \right)\}_{i=1}^{n}  $ be an i.i.d.\ sample, which is distributed according to some probability distribution $P$, that is, the measure $P(G)=\mathbb{P}((Y_i,D_i,\boldsymbol{Z}_i)\in G)$ for all $G\in\mathcal{B}_{\mathbb{R}^{2+L}}$. 
	For every measurable function $v$, by an abuse of notation, we define
	\begin{align}\label{eq.Q map}
		P\left(  v\right)  =\int v\,\mathrm{d}P.
	\end{align}
    For every $\left(  h,g\right)  \in {\bar{\mathcal{H}}\times\mathcal{G}}$ with $g=(g_{1},g_{2})$, define
	\begin{align}\label{eq.phi_Q}
		\phi\left(  h,g\right)    =\frac{P\left(  h\cdot g_{2}\right)
		}{P\left(  g_{2}\right)  }-\frac{P\left(  h\cdot g_{1}\right)  }{P\left(
			g_{1}\right)  }.
	\end{align}
   The null hypothesis equivalent to \eqref{eq.testable implication inequalities multi} is 
	\begin{align}\label{eq.null order}
	\mathrm	H_0: \sup_{\left(  h,g\right)  \in
			{{\mathcal{H}}\times\mathcal{ G}}}\phi\left( h,g\right)\le 0,
	\end{align}
	and the alternative hypothesis is 
	\begin{align*}
	\mathrm	H_1:\sup_{\left(  h,g\right)  \in
			{{\mathcal{H}}\times\mathcal{ G}}}\phi\left( h,g\right)> 0.
	\end{align*} 
	We then define the sample analogue of $\phi$ by
	\begin{align*}
		\hat{\phi}\left(  h,g\right)  =\frac{\hat{P}(h\cdot g_{2})  }{\hat{P}(g_{2})}-\frac{\hat{P}(h\cdot g_{1}) }{\hat{P}(g_{1}) },
	\end{align*}
	where $\hat{P}$ denotes the empirical probability measure of $P$ such that for every measurable function $v$,
	\begin{align}\label{eq.defPn order}
		\hat{P}\left(  v\right)  =\frac{1}{n}\sum_{i=1}^{n}v\left(  Y_{i},D_{i},\boldsymbol{Z}_{i}\right),
	\end{align}
	and $\{(  Y_{i},D_{i},\boldsymbol{Z}_{i})\}_{i=1}^n$ is the i.i.d.\ sample distributed according to $P$. Define 
 \begin{align}\label{eq.estimated stat variance multi order}
		\hat{\sigma}^{2}\left(  h,g\right)  =\frac{T_n}{n}\cdot\left\{\frac{ \hat{P}\left(  h^2\cdot g_{2}\right)   }{\hat{P}^{2}\left(
			g_{2}\right)  }  -\frac{ \hat{P}^2\left(  h\cdot g_{2}\right)   }{\hat{P}^{3}\left(
			g_{2}\right)  }  
		+\frac{ \hat{P}\left(	h^2\cdot g_{1}\right)   }{\hat{P}^{2}\left(  g_{1}\right)  }
		-\frac{ \hat{P}^2\left(	h\cdot g_{1}\right)   }{\hat{P}^{3}\left(  g_{1}\right)  }\right\}
	\end{align}
 for all		$(h,g)\in\bar{\mathcal{H}}\times\mathcal{G}$ with $g=(g_1,g_2)$, where $T_n=n\cdot\prod_{k=1}^{K}\hat{P}\left(1_{\mathbb{R}\times\mathbb{R}\times\{\boldsymbol{z}_k\}}\right)$.


	We then specify a closed set $\Xi\subset(0,1]$ such that $\Xi$ contains all the values of $\xi$ used for constructing the test statistic in the following.  
	We also specify a positive measure $\nu$ on $\Xi$ that satisfies the following assumption. 
	\begin{assumption}\citep{sun2018ivvalidity}\label{ass.nu order}
		The measure $\nu$ satisfies that $0<\nu(\Xi)<\infty$ and ${S}_n\in L^{1}(\nu)$ for all $\omega\in\Omega$ and all $n$ with
		\begin{align*}
			{S}_n(\xi)= 	\sup_{(h,g)\in{\mathcal{H}}\times\mathcal{G}} \frac{\hat{\phi}(h,g)}{\max\{\xi,\hat{\sigma}(h,g)\}}.
		\end{align*}
	\end{assumption}
	We follow \citet{kitagawa2015test} and \citet{sun2018ivvalidity} and construct the test statistic as
	\begin{equation}\label{eq.test stat expansion order}
		TS_n=\int_{\Xi}\sup_{(h,g)\in{\mathcal{H}}\times\mathcal{G}} \frac{\sqrt{T_n}\hat{\phi}(h,g)}{\max\{\xi,\hat{\sigma}(h,g)\}}\,\mathrm{d}\nu(\xi).
	\end{equation}
	We may set the measure $\nu$ to be a Dirac measure centered at some fixed $\xi\in\Xi$, and this is equivalent to using a particular value of $\xi$ to construct the statistic. 
	We construct a random set $\widehat{\Psi_{{\mathcal{H}}\times\mathcal{G}}}$  by
	\begin{align}\label{eq.Psi_hat order}
		\widehat{\Psi_{{\mathcal{H}}\times\mathcal{G}}}=\left\{  \left(  h,g\right)
		\in{\mathcal{H}}\times\mathcal{G}:\sqrt{T_n}\left\vert \frac{\hat{\phi}(h,g)}{\max\{\xi_0,\hat{\sigma}(h,g)\}}\right\vert  \leq\tau
		_{n}\right\}
	\end{align}
	with $\tau_{n}\rightarrow\infty$ and $\tau_{n}/\sqrt{n}\rightarrow0$ as
	$n\rightarrow\infty$, where $\xi_0$ is a fixed small positive number. In practice, we suggest setting $\xi_0=10^{-10}$ which is used in the simulations and the application in the paper. 
	This random set is an estimator for some contact set similar to those in \citet{Beare2015improved} and  \citet{Beare2017improved} in different contexts.\footnote{See \citet{linton2010improved} and \citet{lee2013testing} for more discussions on estimation of contact sets.}  Algorithm \ref{proc:test} illustrates the test procedure. 
 
\begin{algorithm}
\caption{Test Procedure}\label{proc:test}	
	\begin{enumerate}[label=(\arabic*)]
		\item Draw a bootstrap sample $\{  (  \hat{Y}_{i},\hat{D}_{i},\hat{\boldsymbol{Z}}_{i})  \}  _{i=1}^{n}$ independently with replacement from the
		sample $\left\{  \left(  Y_{i},D_{i},\boldsymbol{Z}_{i}\right)  \right\}  _{i=1}^{n}$.
		
		\item Let $\hat{P}^{*}\left(  v\right)  =n^{-1}\sum_{i=1}^{n}v(  \hat{Y}_{i},\hat{D}_{i},\hat{\boldsymbol{Z}}_{i})  $ for all measurable function $v$, and $T_n^{*}=n\cdot\prod_{k=1}^{K}\hat{P}^{*}(1_{\mathbb{R}\times\mathbb{R}\times\{\boldsymbol{z}_k\}})$. Construct the bootstrap version of $\hat{\phi}$ by
		\begin{align*}\label{eq.phi hat star order}
			\hat{\phi}^{*}\left(  h,g\right)  =\frac{\hat{P}^{*}\left(  h\cdot
				g_{2}\right)  }{\hat{P}^{*}\left(  g_{2}\right)  }-\frac{\hat{P}^{*}\left(
				h\cdot g_{1}\right)  }{\hat{P}^{*}\left(  g_{1}\right)  },
		\end{align*}
		 and the bootstrap version of $\hat{\sigma}$ by
		\begin{align*}
			\hat{\sigma}^{*}\left(  h,g\right)  =\sqrt{\frac{T_n^{*}}{n}}\cdot\sqrt{\frac{ \hat{P}^{*}\left(  h^2\cdot g_{2}\right)   }{\hat{P}^{*}\left(g_{2}\right)^2  }  -\frac{ \hat{P}^{*}\left(  h\cdot g_{2}\right)^2   }{\hat{P}^{*}\left(
					g_{2}\right)^3  }  
				+\frac{ \hat{P}^{*}\left(	h^2\cdot g_{1}\right)   }{\hat{P}^{*}\left(  g_{1}\right)^2  }
				-\frac{ \hat{P}^{*}\left(	h\cdot g_{1}\right)^2   }{\hat{P}^{*}\left(  g_{1}\right)^3  }  }
		\end{align*}
		for all
		$(h,g)\in\bar{\mathcal{H}}\times\mathcal{G}$ with $g=(g_1,g_2)$.
		
		\item Construct the bootstrap version of the test statistic by  
		\begin{align*}
			TS_n^{*}=\int_{\Xi}\sup_{(h,g)\in{\widehat{\Psi_{{\mathcal{H}}\times\mathcal{G}}}}} \frac{\sqrt{T_n^{*}}(  \hat{\phi}^{*}(h,g)-\hat{\phi}(h,g))}{\max\{\xi,\hat{\sigma}^{*}(h,g)\}}\,\mathrm{d}\nu(\xi).
		\end{align*}
		
		\item Repeat steps (1), (2), and (3) $n_B$ times independently with $n_B$ chosen as large as is computationally convenient. Given
	a prespecified nominal significance level $\alpha$, we construct the bootstrap critical value
		$\hat{c}_{1-\alpha}$ by
		\begin{align*}
			\hat{c}_{1-\alpha}=\inf\left\{  c:\mathbb{P}\left(  TS_n^{*}  \leq c\bigg|\{(Y_{i},D_{i},Z_{i})\}_{i=1}^{n}\right)  \geq
			1-\alpha\right\}  .
		\end{align*}
		In practice, we may approximate $\hat{c}_{1-\alpha}$ by the $1-\alpha$ quantile of the $n_B$ independently generated bootstrap statistics.

		\item The decision rule for the test is: Reject $\mathrm H_{0}$ if $TS_n>\hat{c}_{1-\alpha}$.

	\end{enumerate}
\end{algorithm}

 \begin{proposition}
		\label{prop.test multi order}Suppose Assumption \ref{ass.nu order} holds. 
		\begin{enumerate}[label=(\roman{*})]
			\item If the $\mathrm H_0$ in \eqref{eq.null order} is true
			and the CDF of the asymptotic limit of $TS_n$ is increasing and continuous at its $1-\alpha$ quantile, then
			$\lim_{n\rightarrow\infty}\mathbb{P}(TS_n  >\hat{c}_{1-\alpha}) =\alpha$.

			\item If the $\mathrm H_0$ in \eqref{eq.null order} is false, then 
			$\lim_{n\rightarrow\infty}\mathbb{P}(
			TS_n  >\hat{c}_{1-\alpha})  =1$.
			
		\end{enumerate}
	\end{proposition} 

 Proposition \ref{prop.test multi order} may be proved analogously to Theorem 3.2 in \citet{sun2018ivvalidity}, so we omit the proof. The simulation study in Section \ref{sec.simulation} in the supplementary appendix demonstrates the good finite sample properties of the test. The numerical results show that the test is asymptotically size controlled and consistent. 
 Section \ref{sec.application} in the appendix provides an empirical application \citep{thornton2008demand} for the proposed test. We examine the partial validity of monetary incentives and distance from results centers as instruments for the knowledge of HIV status, and find that these instruments passed our test. 
	\bibliographystyle{apalike}
	\bibliography{reference1}


\newpage 
\appendix

\setcounter{page}{1} 

\begin{center}
    \huge{Testing Partial Instrument Monotonicity}
 
    \Large{Online Supplementary Appendix}
    
    \vspace{0.5cm}
    
    \large{Hongyi Jiang \qquad Zhenting Sun}
\end{center}

\startcontents[sections]
\printcontents[sections]{l}{1}{\setcounter{tocdepth}{2}}

	\section{Simulation Study} \label{sec.simulation}

 We design simulations based on those in \citet{kitagawa2015test} and \citet{sun2018ivvalidity}. The key difference is that our simulations are for partial monotonicity of multi-dimensional instruments. 
	Throughout this section, we set $D\in\{0,1,2\}$ and $\boldsymbol{Z}=(Z_1,Z_2)\in\{(0,0),(0,1),(1,0),(1,1)\}$. For each
	simulation, we set the Monte Carlo iteration to $1000$ and the bootstrap iteration to $1000$. Since the computational workload is heavy, the warp-speed
	method of \citet{giacomini2013warp} is employed in the simulations. 
	We choose the measure $\nu$ as a Dirac measure $\delta_{\xi}$ centered at one of the following values of $\xi$: $0.02$, $0.03$, $0.04$, $0.05$, $0.06$, $0.07$, $0.08$, $0.09$, $0.1$, and $1$, or as a probability measure $\bar{\nu}_{\xi}$ that assigns equal probabilities (weights) to the values of $\xi$ listed above.  For all the simulations, we set the nominal significance level $\alpha$ to $0.05$. 
	
	\subsection{Size Control and Tuning Parameter Selection}\label{sec.size}
	We first run simulations to show the size control of the test. As shown in \eqref{eq.Psi_hat order}, a tuning parameter $\tau_n$ needs to be specified for
 the estimate $\widehat{\Psi_{{\mathcal{H}}\times\mathcal{G}}}$ for every $n$. In this section, we set $n$ to $2000$ and $\tau_n$ to $0.1$, $0.5$, $1$, $2$, $3$, $4$, and $\infty$. As shown in \citet{sun2018ivvalidity}, for $\tau_n=\infty$, $\widehat{\Psi 		_{{\mathcal{H}}\times\mathcal{G}}}={{\mathcal{H}}	\times\mathcal{G}}$ and the test is conservative. 
 We then compare the rejection rates obtained using each of these values of $\tau_n$ and choose its value for sample sizes close to $2000$. 
 
 We let $U_1\sim\mathrm{Unif}(0,1)$, $U_2\sim\mathrm{Unif}(0,1)$, $V\sim\mathrm{Unif}(0,1)$, $N_0\sim \mathrm{N}(0,1)$, $N_1\sim \mathrm{N}(1,1)$, $N_2\sim \mathrm{N}(2,1)$, $Z_l=1\{U_l \le 0.5\}$  ($\mathbb{P}(Z_l=1)=0.5$) for $l=1,2$, $D_{\boldsymbol{z}}=2\times 1\{V\le 0.33\}+ 1\{0.33  <V\le 0.66\}$ for $\boldsymbol{z}\in\{(0,0),(0,1),(1,0),(1,1)\}$,  $D=\sum_{\boldsymbol{z}} 1\{\boldsymbol{Z}=\boldsymbol{z}\}\times D_{\boldsymbol{z}}$, and $Y=\sum_{d=0}^{2}1\{D=d\}\times N_d$. All the variables $U_1$, $U_2$, $V$, $N_0$, $N_1$, and $N_2$ are mutually independent. Clearly, Assumption \ref{ass.IV validity for multivalued Z}(iii) holds in this case for $\{(0,0),(0,1)\}$, $\{(1,0),(1,1)\}$, $\{(0,0),(1,0)\}$, and $\{(0,1),(1,1)\}$.
	
	Table \ref{tab:Rej H0 multi} shows the simulation results. As expected, for each measure $\nu$, a smaller $\tau_n$ yields greater rejection rates. This is because a smaller $\tau_n$ produces a smaller critical value according to \eqref{eq.Psi_hat order}. 
	For $\tau_n=2$, all the rejection rates are close to those for $\tau_n=\infty$ (the conservative case) and close to the significance level $\alpha$. Thus when sample sizes are less than or close to $2000$, we suggest choosing $\tau_n=2$ in practice to achieve good size control with relatively high empirical powers.  Clearly, when the sample size increases, $\tau_n$ should be increased accordingly. Additional simulations can be conducted to obtain $\tau_n$ for larger sample sizes. 
 
	\begin{table}[h]
		
		\centering
		\caption{Rejection Rates under $\mathrm H_{0}$ for Multi-dimensional $Z$}
		\scalebox{0.9}{
			\begin{tabular}{   c  c  c  c  c  c  c  c  c  c  c  c  }
				\hline
				\hline
				\multirow{2}{*}{$\tau_n$} & \multicolumn{10}{c}{$\xi$ for $\delta_{\xi}$} &\multirow{2}{*}{$\bar{\nu}_{\xi}$} \\
				\cline{2-11}
				& 0.02 & 0.03 & 0.04 & 0.05 & 0.06 & 0.07 & 0.08 & 0.09 & 0.1 & 1 & \\
				\hline
				
				0.1 & 0.139 & 0.112 & 0.098 & 0.090 & 0.101 & 0.095 & 0.091 & 0.088 & 0.088 & 0.088 & 0.103 \\ 
				0.5 & 0.083 & 0.071 & 0.072 & 0.046 & 0.055 & 0.059 & 0.060 & 0.059 & 0.059 & 0.059 & 0.055 \\ 
				1 & 0.056 & 0.059 & 0.058 & 0.039 & 0.049 & 0.052 & 0.054 & 0.053 & 0.053 & 0.053 & 0.047 \\ 
				2 & 0.040 & 0.055 & 0.049 & 0.038 & 0.034 & 0.035 & 0.049 & 0.040 & 0.040 & 0.040 & 0.037 \\ 
				3 & 0.039 & 0.055 & 0.048 & 0.038 & 0.034 & 0.035 & 0.044 & 0.040 & 0.040 & 0.040 & 0.037 \\ 
				4 & 0.039 & 0.055 & 0.048 & 0.038 & 0.034 & 0.035 & 0.044 & 0.040 & 0.040 & 0.040 & 0.037 \\ 
				$\infty$ & 0.039 & 0.055 & 0.048 & 0.038 & 0.034 & 0.035 & 0.044 & 0.040 & 0.040 & 0.040 & 0.037 \\

				\hline\hline
			\end{tabular}
		}
		
		\label{tab:Rej H0 multi}
	\end{table}
	
	\subsection{Empirical Power}
	In this section, we run simulations to investigate the power of the test. We design six data generating processes (DGPs) for all of which Assumption \ref{ass.IV validity for multivalued Z} does not hold. We consider five sample sizes $n=200$, $600$, $1000$, $1100$, and $2000$, and set the probability $\mathbb{P}(Z_j=1)=r_n$ for $j=1,2$, with $r_n=1/2$, $1/6$, $1/2$, $1/11$, and $1/2$ for the sample sizes listed above. We set $\tau_n=2$ as selected in Section \ref{sec.size}. 
 
    We let $U_1\sim\mathrm{Unif}(0,1)$, $U_2\sim\mathrm{Unif}(0,1)$, $V\sim\mathrm{Unif}(0,1)$, $W\sim\mathrm{Unif}(0,1)$, and $Z_j=1\{U_j \le r_n\}$ for $j=1,2$.
	For DGPs (1)--(4), we let $D_{\boldsymbol{z}}=2\times 1\{V\le 0.45\}+ 1\{0.45 <V\le 0.55\}$ for every $\boldsymbol{z}\in\{(0,0),(0,1),(1,0),(1,1)\}$,   $D=\sum_{\boldsymbol{z}} 1\{\boldsymbol{Z}=\boldsymbol{z}\}\times D_{\boldsymbol{z}}$, $N_{0,(0,0)}\sim \mathrm{N}(0,1)$, $N_{1,(0,0)}\sim \mathrm{N}(0,1)$, and $N_{d \boldsymbol{z}}\sim \mathrm{N}(0,1)$ for $d=0,1,2$ and $\boldsymbol{z}\in\{(0,1),(1,0),(1,1)\}$.
	\begin{enumerate}[label=(\arabic*):]
		
		\item  $N_{2,(0,0)}\sim \mathrm{N}(-0.7,1)$ and $Y=\sum_{\boldsymbol{z}}1\{\boldsymbol{Z}=\boldsymbol{z}\}\times(\sum_{d=0}^{2} 1\{D=d\}\times N_{d\boldsymbol{z}})$.
		
		\item $N_{2,(0,0)}\sim \mathrm{N}(0,1.675^2)$ and $Y=\sum_{\boldsymbol{z}}1\{\boldsymbol{Z}=\boldsymbol{z}\}\times(\sum_{d=0}^{2} 1\{D=d\}\times N_{d\boldsymbol{z}})$.
		
		\item $N_{2,(0,0)}\sim \mathrm{N}(0,0.515^2)$ and $Y=\sum_{\boldsymbol{z}}1\{\boldsymbol{Z}=\boldsymbol{z}\}\times(\sum_{d=0}^{2} 1\{D=d\}\times N_{d\boldsymbol{z}})$.
		
		\item $N_{2,(0,0)a}\sim \mathrm{N}(-1,0.125^2)$, $N_{2,(0,0)b}\sim \mathrm{N}(-0.5,0.125^2)$, $N_{2,(0,0)c}\sim \mathrm{N}(0,0.125^2)$,\\ $N_{2,(0,0)d}\sim \mathrm{N}(0.5,0.125^2)$, $N_{2,(0,0)e}\sim \mathrm{N}(1,0.125^2)$, $N_{2,(0,0)}=1\{W\le 0.15\}\times N_{2,(0,0)a}+1\{0.15<W\le 0.35\}\times N_{2,(0,0)b}+1\{0.35<W\le 0.65\}\times N_{2,(0,0)c}+1\{0.65<W\le 0.85\}\times N_{2,(0,0)d}+1\{W>0.85\}\times N_{2,(0,0)e}$, and $Y=\sum_{\boldsymbol{z}}1\{\boldsymbol{Z}=\boldsymbol{z}\}\times(\sum_{d=0}^{2} 1\{D=d\}\times N_{d\boldsymbol{z}})$.
		
	\end{enumerate} 
	For DGPs (5) and (6), we let $N_0\sim \mathrm{N}(0,1)$, $N_1\sim \mathrm{N}(1,1)$, and $N_2\sim \mathrm{N}(2,1)$.
	\begin{enumerate}[resume,label=(\arabic*):]
		
		\item $D_{(0,0)}=2\times 1\{V\le 0.6\}+ 1\{0.6 <V\le 0.8\}$, $D_{(0,1)}=2\times 1\{V\le 0.33\}+ 1\{0.33 <V\le 0.66\}$, $D_{(1,0)}=D_{(1,1)}=D_{(0,1)}$, $D=\sum_{\boldsymbol{z}} 1\{\boldsymbol{Z}=\boldsymbol{z}\}\times D_{\boldsymbol{z}}$, and $Y=\sum_{d=0}^{2} 1\{D=d\}\times N_{d}$.
		
		\item $D_{(0,0)}=2\times 1\{V\le 0.33\}+ 1\{0.33 <V\le 0.66\}$, $D_{(0,1)}=2\times 1\{V\le 0.6\}+ 1\{0.6 <V\le 0.8\}$, $D_{(1,0)}=D_{(1,1)}=D_{(0,0)}$, $D=\sum_{\boldsymbol{z}} 1\{\boldsymbol{Z}=\boldsymbol{z}\}\times D_{\boldsymbol{z}}$, and $Y=\sum_{d=0}^{2} 1\{D=d\}\times N_{d}$.
		
	\end{enumerate}
	
	All the variables $U_1$, $U_2$, $V$, $N_{0,(0,0)}$, $N_{1,(0,0)}$, $N_{2,(0,0)}$, $N_{0,(1,0)}$, $N_{1,(1,0)}$, $N_{2,(1,0)}$, $N_{0,(0,1)}$, $N_{1,(0,1)}$, $N_{2,(0,1)}$, $N_{0,(1,1)}$, $N_{1,(1,1)}$, $N_{2,(1,1)}$, $N_0$, $N_1$, and $N_2$ are set to be mutually independent for each DGP. 
	
	Table \ref{tab:Rej H1 multi} shows the simulation results under DGPs (1)--(6). The rejection rate approaches one as the sample size $n$ increases for each DGP and each measure $\nu$.   
	
	\begin{table}[h]
		
		\centering
		\caption{Rejection Rates under $\mathrm H_{1}$ for Multivalued $D$ and Multivalued $Z$}
		\scalebox{0.85}{
			\begin{tabular}{  c  c  c  c  c  c  c  c  c  c  c  c  c  }
				\hline
				\hline
				\multirow{2}{*}{DGP} & \multirow{2}{*}{$n$}& \multicolumn{10}{c}{$\xi$ for $\delta_{\xi}$} &\multirow{2}{*}{$\bar{\nu}_{\xi}$} \\
				\cline{3-12}
				&  & 0.02 & 0.03 & 0.04 & 0.05 & 0.06 & 0.07 & 0.08 & 0.09 & 0.1 & 1 & \\
				\hline

				\multirow{5}{*}{(1)}	
				& 200 & 0.028 & 0.096 & 0.123 & 0.137 & 0.146 & 0.143 & 0.110 & 0.089 & 0.089 & 0.089 & 0.124 \\ 
				& 600 & 0.074 & 0.059 & 0.053 & 0.053 & 0.047 & 0.044 & 0.043 & 0.043 & 0.043 & 0.043 & 0.057 \\ 
				& 1000 & 0.706 & 0.746 & 0.775 & 0.821 & 0.797 & 0.769 & 0.700 & 0.649 & 0.648 & 0.648 & 0.802 \\ 
				& 1100 & 0.041 & 0.046 & 0.064 & 0.049 & 0.049 & 0.049 & 0.049 & 0.049 & 0.049 & 0.049 & 0.050 \\ 
				& 2000 & 0.991 & 0.996 & 0.998 & 0.997 & 0.997 & 0.995 & 0.988 & 0.988 & 0.988 & 0.988 & 0.998 \\ 		
				\hline		
				
				\multirow{5}{*}{(2)}	
				
				& 200 & 0.004 & 0.031 & 0.045 & 0.052 & 0.055 & 0.052 & 0.043 & 0.038 & 0.038 & 0.038 & 0.047 \\ 
				& 600 & 0.051 & 0.050 & 0.051 & 0.053 & 0.047 & 0.044 & 0.043 & 0.043 & 0.043 & 0.043 & 0.053 \\ 
				& 1000 & 0.366 & 0.436 & 0.391 & 0.265 & 0.130 & 0.071 & 0.049 & 0.042 & 0.042 & 0.042 & 0.162 \\ 
				& 1100 & 0.041 & 0.046 & 0.064 & 0.049 & 0.049 & 0.049 & 0.049 & 0.049 & 0.049 & 0.049 & 0.049 \\ 
				& 2000 & 0.952 & 0.959 & 0.906 & 0.773 & 0.458 & 0.263 & 0.137 & 0.126 & 0.126 & 0.126 & 0.697 \\ 
				
				\hline
				
				\multirow{5}{*}{(3)}	
				
				& 200 & 0.036 & 0.144 & 0.181 & 0.194 & 0.208 & 0.231 & 0.209 & 0.183 & 0.183 & 0.183 & 0.199 \\ 
				& 600 & 0.129 & 0.103 & 0.064 & 0.060 & 0.047 & 0.044 & 0.043 & 0.043 & 0.043 & 0.043 & 0.074 \\ 
				& 1000 & 0.806 & 0.842 & 0.867 & 0.890 & 0.907 & 0.908 & 0.887 & 0.867 & 0.867 & 0.867 & 0.902 \\ 
				& 1100 & 0.041 & 0.046 & 0.064 & 0.049 & 0.049 & 0.049 & 0.049 & 0.049 & 0.049 & 0.049 & 0.050 \\ 
				& 2000 & 0.984 & 0.994 & 0.995 & 0.996 & 0.997 & 0.999 & 0.998 & 0.998 & 0.998 & 0.998 & 0.998 \\ 
				
				\hline
				
				\multirow{5}{*}{(4)}	
				& 200 & 0.015 & 0.057 & 0.083 & 0.114 & 0.108 & 0.119 & 0.111 & 0.103 & 0.103 & 0.103 & 0.103 \\ 
				& 600 & 0.069 & 0.074 & 0.076 & 0.080 & 0.080 & 0.075 & 0.059 & 0.059 & 0.059 & 0.059 & 0.078 \\ 
				& 1000 & 0.298 & 0.371 & 0.431 & 0.428 & 0.468 & 0.491 & 0.507 & 0.495 & 0.495 & 0.495 & 0.498 \\ 
				& 1100 & 0.045 & 0.053 & 0.068 & 0.056 & 0.049 & 0.049 & 0.049 & 0.049 & 0.049 & 0.049 & 0.056 \\ 
				& 2000 & 0.748 & 0.790 & 0.777 & 0.784 & 0.829 & 0.849 & 0.859 & 0.851 & 0.851 & 0.851 & 0.875 \\

				\hline
				
				\multirow{5}{*}{(5)}	
				& 200 & 0.438 & 0.618 & 0.680 & 0.739 & 0.751 & 0.799 & 0.848 & 0.852 & 0.852 & 0.852 & 0.784 \\ 
				& 600 & 0.678 & 0.770 & 0.639 & 0.541 & 0.435 & 0.365 & 0.352 & 0.352 & 0.352 & 0.352 & 0.641 \\ 
				& 1000 & 1.000 & 1.000 & 1.000 & 1.000 & 1.000 & 1.000 & 1.000 & 1.000 & 1.000 & 1.000 & 1.000 \\ 
				& 1100 & 0.220 & 0.133 & 0.090 & 0.069 & 0.069 & 0.069 & 0.069 & 0.069 & 0.069 & 0.069 & 0.139 \\ 
				& 2000 & 1.000 & 1.000 & 1.000 & 1.000 & 1.000 & 1.000 & 1.000 & 1.000 & 1.000 & 1.000 & 1.000 \\

				\hline
				
				\multirow{5}{*}{(6)}
				& 200 & 0.175 & 0.375 & 0.472 & 0.530 & 0.566 & 0.614 & 0.663 & 0.688 & 0.686 & 0.686 & 0.586 \\ 
				& 600 & 0.308 & 0.369 & 0.492 & 0.551 & 0.540 & 0.505 & 0.494 & 0.494 & 0.494 & 0.494 & 0.524 \\ 
				& 1000 & 0.991 & 0.997 & 0.998 & 1.000 & 1.000 & 1.000 & 1.000 & 1.000 & 1.000 & 1.000 & 1.000 \\ 
				& 1100 & 0.334 & 0.406 & 0.367 & 0.317 & 0.317 & 0.317 & 0.317 & 0.317 & 0.317 & 0.317 & 0.387 \\ 
				& 2000 & 1.000 & 1.000 & 1.000 & 1.000 & 1.000 & 1.000 & 1.000 & 1.000 & 1.000 & 1.000 & 1.000 \\

				\hline\hline
			\end{tabular}
		}
		
		\label{tab:Rej H1 multi}
	\end{table}
		
	\section{Empirical Application}\label{sec.application}
	
		We visit one empirical example discussed by \citet{mogstad2021causal} to show the performance of the proposed test in practice. The example is from \citet{thornton2008demand}, who evaluates an experiment in rural Malawi where individuals who were screened for HIV were then randomly assigned incentives to receive their results. 
  \citet{thornton2008demand} estimates the causal effect of learning HIV status on the demand for condoms using
  a cash-redeemable voucher and the distance to the nearest results center as instruments for the decision to obtain the results.
    In this example, \citet{mogstad2021causal} claim that the partial instrument monotonicity is likely to hold: Given the distance fixed, if an individual gets a voucher incentive, she is more likely to obtain her results. Given the voucher incentive, the opportunity cost will decrease if an individual is randomly assigned a closer results facility, and therefore the individual is more likely to obtain the test results.
    As described by \citet{thornton2008demand}, the dataset consists of 3 sub-datasets: The 2004 baseline data collected by MDICP4 for men, women, boys, and girls; the randomly assigned locations of VCT clinic which create the distance to the clinic variables; the 2005 follow-up data collected by \citet{thornton2008demand} as the dissertation field work.
	
	In this dataset, the available sample size is 1528. The treatment $D$ is a dummy variable for obtaining HIV results (the variable ``got''), the instrument $\boldsymbol{Z}=(Z_1,Z_2)$ is 2-dimensional, where $Z_1$ is a dummy for receiving any positive valued incentive (the variable ``any'') and $Z_2$ is a dummy for living under 1.5 km from the assigned VCT center (the variable ``under''), and the outcome $Y$ is the number of condoms purchased at the follow‐up survey (the variable ``numcond'') in the dataset. 
	{Table} \ref{tab:ApplicationThornton} shows the $p$-values obtained from our test using each measure $\nu$ with $\tau_n=2$, which suggest that we do not reject the partial validity of instrument $\boldsymbol{Z}$ as expected by \citet{mogstad2021causal}.
	
	\begin{table}[h]
		
		\centering
		\caption{$p$-values Obtained from the Proposed Test for Each Measure $\nu$}
		\scalebox{0.9}{
			\begin{tabular}{ c  c  c  c  c  c  c  c  c  c c}
				\hline
				\hline
				\multicolumn{10}{c}{$\xi$ for $\delta_{\xi}$} &\multirow{2}{*}{$\bar{\nu}_{\xi}$} \\
				\cline{1-10}
				0.02 & 0.03 & 0.04 & 0.05 & 0.06 & 0.07 & 0.08 & 0.09 & 0.1 &1& \\
				\hline
				
				0.846&	0.726&	0.616&	0.491&	0.482&	0.503&	0.533&	0.548&	0.548&	0.548&	0.619

				\\				
				
				\hline\hline
			\end{tabular}
		}
		
		\label{tab:ApplicationThornton}
	\end{table}

\end{document}